\documentclass[assymb,11pt]{article}
\usepackage[pdftex,colorlinks=true,urlcolor=black,filecolor=black,linkcolor=black,
            pdftitle={Constructing Spacetime.},
            pdfauthor={Mark D. Roberts},
            pdfsubject={relativity,geometry},
            pdfkeywords={finsler,lagrange},
            pagebackref,pdfpagemode=None,bookmarksopen=true]{hyperref}
\usepackage{amssymb}
\newcommand{\bc}{\begin{center}}
\newcommand{\ec}{\end{center}}
\newcommand{\bi}{\begin{itemize}}     
\newcommand{\ei}{\end{itemize}}
\newcommand{\bd}{\begin{description}} 
\newcommand{\ed}{\end{description}}
\newcommand{\bn}{\begin{enumerate}}   
\newcommand{\en}{\end{enumerate}}
\newcommand{\be}{\begin{equation}}
\newcommand{\ee}{\end{equation}}
\newcommand{\ber}{\begin{eqnarray}}
\newcommand{\ear}{\end{eqnarray}}
\newcommand{\ba}{\begin{array}}
\newcommand{\ea}{\end{array}}
\newcommand{\bv}{\begin{verbatim}}
\newcommand{\ev}{\end{verbatim}}
\newcommand{\al}{\alpha}
\newcommand{\ar}{{\cal A}}

\newcommand{\ds}{\rm{d}\sigma}
\newcommand{\dt}{\rm{d}\tau}

\newcommand{\el}{\ell}

\newcommand{\ff}{\aleph}
\newcommand{\fr}{\frac}
\newcommand{\Ga}{\Gamma}
\newcommand{\ga}{\gamma}
\newcommand{\gm}{\hat{g}}

\newcommand{\lb}{\label}

\newcommand{\Lg}{{\cal L}}
\newcommand{\n}{\nonumber\\}

\newcommand{\si}{\sigma}
\newcommand{\sq}{\sqrt}
\newcommand{\st}{\stackrel}
\newcommand{\ta}{\tau}

\newcommand{\p}{\partial}
\begin{document}
\title{Constructing Spacetime from the String Worldsheet.}
\author{
\href{http://www.violinist.com/directory/bio.cfm?member=robemark}
{Mark D. Roberts},\\
54 Grantley Avenue,  Wonersh Park,  GU5 0QN,  UK\\
mdr@ihes.fr
}
\date{$2^{nd}$ of November 2009}
\maketitle
\begin{abstract}
In a certain sense riemannian geometry can be thought of as geometry built up from the
finslerian properties of point particles.
The generalization of this to where the geometry is built up from the finslerian properties
of string and membranes is investigated.
Solely classical arguments suggest a physical interpretation in
which microscopic strings are directly related to macroscopic geometry;
alternatively the resulting geometry can be interpretated as that describing
microsopic spacetime.
\end{abstract}
\vspace{1cm}
{\small\tableofcontents}
\section{Introduction}
A problem with string theory is the nature of the relationship
between microscopic strings and macroscopic spacetime.
A property of string theory is that, in the limit that the string becomes a point,
a corresponding field theory is recovered.
General relativity is based upon geometry defined at points:
from a finslerian point of view the riemmanian geometry of general relativity
is just a macroscopic geometry constructed from the point particles' lagrangian.
This leads to the question of what sort of geometry corresponds
to the string and membrane lagrangians.
This is the problem looked at here.
Geometry based upon area,  although not world sheet area,  has been studies by
Cartan \cite{cartan}, see also Akivis and Rosenfeld \cite{AR},
Brickell \cite{brickell} and Vacaru \cite{vacaru};
Brickell considers dependence on position and area.
The difference of these approaches to the present one is that
dependence on velocities is also needed.
In modern canonical gravity objects which are not point-like
are sometimes studied,  see for example Abdalla, Castello-Branco and Lima-Santos \cite{ACL}.
Spacetime as an emergent structure has been studied by
Barcelo, Liberati and Visser \cite{BLV}, Bekenstein \cite{bekenstein} and Pavsic \cite{pavsic}.
Textbooks on finslerian geometry include Bejancu \cite{bejancu},
Matsumato \cite{matsumato} and Rund \cite{rund},
here Rund Chap.1 is followed where possible.
The terminology used here is that $\gm$ is the hat metric
and $g$ is the indicial metric used for simple raising and loweing of indices.
The systems that Rund \cite{rund} considers have hamiltonian normalized to one,
as here systems which are weakly zero are often used Rund's hamiltonian
approach is not gone into.
The connection and curvature can be defined several ways,
here we use the familar Christoffel connection and Riemann curvature except that now
the metric can be velocity dependent.
\newpage
\section{Finsler geometry.}
Following Rund \cite{rund} pages 1-23
assume that a given space has points, curves, and velocities
\be
x^i,~~~~~x^i(t),~~~~~\dot{x}^i\equiv\fr{d x^i(t)}{dt},~~~~~\sum_i(\dot{x}^i)\ne0.
\label{r1.1r1.3r1.4r1.16b}
\ee
The distance between two close points $A(x^i)$ and $B(x^i+dx^i)$ is given by
\be
ds=F(x^i,dx^j).
\label{r1.14}
\ee
Introduce the notation
\be
F_{\dot{x}^i}(x,\dot{x})\equiv\fr{\p}{\p\dot{x}^i}F(x,\dot{x}).
\label{norund}
\ee
The function $F(x^i,\dot{x}^i)$ is positively homogeneous of degree one in the $\dot{x}^i$
\be
F(x^i,k\dot{x}^i)=kF(x^i,\dot{x}^i),~~~k>0.
\label{r1.8}
\ee
Euler's theorem on homogeneous functions can be expressed as
\be
F_{\dot{x}^i}(x,\dot{x})\dot{x}^i=F(x,\dot{x}),~~~
F_{\dot{x}^i\dot{x}^j}(x,\dot{x})\dot{x}^i=0,~~~
\det\left|F_{\dot{x}^i\dot{x}^j}\right|=0.
\label{r1.11r1.12r1.13}
\ee
Using the notation
\be
F\equiv F(x,\dot{x}),~~~~~~~
F^2_{\dot{x}^i}\equiv\left(F(x,\dot{x})^2\right)_{\dot{x}^i},
\label{notF}
\ee
the chain rule is
\be
\frac{1}{2}F^2_{\dot{x}^i\dot{x}^j}=F_{\dot{x}^i}F_{\dot{x}^j}+FF_{\dot{x}^i\dot{x}^j},
\label{chain}
\ee
multipling by $\dot{x}^i\dot{x}^j$ and applying Euler's theorem \ref{r1.11r1.12r1.13} gives
the second order equality
\be
\fr{1}{2}F^2_{\dot{x}^i\dot{x}^j}(x,\dot{x})\dot{x}^i\dot{x}^j=F^2(x,\dot{x}).
\label{r1.19b}
\ee
One can define a hat metric
\be
\gm_{ij}(x,\dot{x})\equiv\fr{1}{2}F^2_{\dot{x}^i\dot{x}^j}(x,\dot{x}),
\label{r3.1}
\ee
then using the second order equality (\ref{r1.19b})
gives the hat metric finsler function relationship
\be
\dot{x}^i\dot{x}^j\gm_{ij}(x,\dot{x})=F^2(x,\dot{x}),
\label{r3.2}
\ee
comparing with (\ref{r1.14}) the familar expression for infinitesimal distance is recovered.
Rund \cite{rund} goes on to discuss hamiltonian systems.
\section{The point particle.}\label{tpp}
For the point particle the non linear or square root form of the
lagrangian is equal to the length
\be
\Lg=F=-m\el\equiv-m\sqrt{-\dot{x}^2}
\label{pplag}
\ee
This lagrangian obeys the homogeneity relatioship \ref{r1.8} as
\be
F(x^i,k\dot{x}^i)=-m\sqrt{-k\dot{x}^a\cdot k\dot{x}^a}=-mk\el=kF.
\label{ppr1.8}
\ee
Euler' homogeneous equations (\ref{r1.11r1.12r1.13}) become
\be
p^i\equiv F_{\dot{x}^i}=\fr{m\dot{x}^i}{\el},~~~
\dot{x}^iF_{\dot{x}^i}=-m\el,~~~
w^{ij}\equiv F_{\dot{x}^i\dot{x}^j}=\fr{m}{\el}h_{ij},~~~
\dot{x}^iF_{\dot{x}^i\dot{x}^j}=0,
\label{pp1.11}
\ee
where the projection operator
\be
h_{ij}\equiv g_{ij}-\fr{\dot{x}^i\dot{x}^j}{\dot{x}^2},
\label{ppproj}
\ee
is the familar one from general relativity,
and $p^i$ and $w^{ij}$ and the momentum and hessain,  compare eq.(12)\cite{mdr33}.
The second order relationship (\ref{r1.19b}) becomes
\be
F^2_{\dot{x}^i}=-2m^2\dot{x}^i,~~~
F^2_{\dot{x}^i\dot{x}^j}=-2m^2g_{ij},~~~
\fr{1}{2}F^2_{\dot{x}^i\dot{x}^j}\dot{x}^j=FF_{\dot{x}^i},~~~
\fr{1}{2}F^2_{\dot{x}^i\dot{x}^j}\dot{x}^i\dot{x}^j=F^2,
\label{ppf2}
\ee
From the second equation (\ref{ppf2}) the hat metric finsler function relationship is given by
\be
\gm_{ij}\equiv-\fr{1}{2m}F^2_{\dot{x}^i\dot{x}^j},
\label{ppm}
\ee
the minus sign coming about because of lorentz signature,
for a positive definite signature there would be a plus sign.
In terms of the momentum and hessian (\ref{pp1.11}),  (\ref{ppm}) is
\be
\gm_{ij}=-\fr{1}{m}p_ip_j+\el w_{ij},
\label{pppw}
\ee
and the hat metric finsler function relationship (\ref{r3.2}) becomes
\be
\dot{x}^i\dot{x}^j\gm_{ij}=-\fr{1}{m}F^2.
\label{ppdf}
\ee
One difference is that from (\ref{r1.11r1.12r1.13})
\be
\det|F_{\dot{x}^i\dot{x}^j}|=\fr{1}{\el}\det|-g|
\label{ppdet}
\ee
It is apparent that $\gm=g$
so that the metric and geometry are the same as that of general relativity.
\section{The string.}\label{string}
For the string the square root or non linear lagrangian is equal to the area
\be
\Lg=F=-\fr{\ar}{2\pi\al'},~~~
\ar\equiv\sqrt{(\dot{x}\cdot x')^2-\dot{x}^2x'^2}.
\label{slag}
\ee
In the velocity only approach $F=F(x^i,\dot{x}^i)$ in particular $F$ is not a function of $x'$.
For the string use $F(x,\dot{x})$ as before and the $x'$ come out as new terms,
there are other possibilities that are discussed under membranes \S\ref{membrane}.
The lagrangian (\ref{slag}) obeys the homogeneity relationship (\ref{r1.8}) as
\be
F(x^i,k\dot{x}^i)=-\fr{1}{2\pi\al'}\sqrt{k^2(\dot{x}\cdot x')^2-k\dot{x}^a\cdot k\dot{x}^ax'^2}
=-\fr{k}{2\pi\al'}\ar=kF,
\label{sr1.8}
\ee
Euler's homogeneous equations (\ref{r1.11r1.12r1.13}) become
\ber
&&p^i\equiv F_{\dot{x}^i}=\fr{1}{2\pi\al'\ar}
\left(-\dot{x}\cdot x'x'^i+x'^2\dot{x}^i\right),~~~~~
\dot{x}^iF_{\dot{x}^i}=\fr{-\ar}{2\pi\al'},\n
&&w^{ij}\equiv F_{\dot{x}^i\dot{x}^j}=
\fr{x'^2h^{ij}}{2\pi\al'\ar},~~~~~
\dot{x}^iF_{\dot{x}^i\dot{x}^j}=0,
\label{st2}
\ear
where the projection tensor is
\be
h^{ij}\equiv g^{ij}+\fr{1}{\ar^2}
\left(\dot{x}^2x'^i x'^j +x'^2\dot{x}^i\dot{x}^j
-(\dot{x}\cdot x')(\dot{x}^i x'^j+x'^i\dot{x}^j)\right),~
h=d-2.
\lb{3.1.16}
\ee
compare \cite{mdr33}\S3.4.
The second order relationship (\ref{r1.19b}) becomes
\ber
&&F^2_{\dot{x}^i}=-\fr{\ar}{\pi\al'}p_i,~~~~~
F^2_{\dot{x}^i\dot{x}^j}=\fr{1}{2\pi^2\al'^2}\left(x'^ix'^j-x'^2g_{ij}\right),\n
&&\fr{1}{2}F^2_{\dot{x}^i\dot{x}^j}\dot{x}^j=FF_{\dot{x}^i},~~~~~
\fr{1}{2}F^2_{\dot{x}^i\dot{x}^j}\dot{x}^i\dot{x}^j=F^2,
\label{stf2}
\ear
The hat metric $\hat{g}$ (\ref{ppm}) generalizes to
\be
\gm_{ij}\equiv-\fr{1}{2}F^2_{\dot{x}^i\dot{x}^j}=\fr{1}{4\pi^2\al'^2}(-x'_ix'_j+x'^2g_{ij}),
\label{stm}
\ee
and in this case does not equal the indicial metric $g$.
In terms of the momentum and hessian (\ref{st2}),  (\ref{stm}) is
\be
\gm_{ij}=-p_ip_j+\fr{\ar}{2\pi\al'}w_{ij},
\label{stpw}
\ee
the hat metric finsler function relationship (\ref{r3.2}) becomes
\be
\dot{x}^i\dot{x}^j\gm_{ij}=-\fr{\ar^2}{4\pi^2\al'^2}=-F^2.
\label{stdf}
\ee
In the present case equation (\ref{stm}) shows that the two metrics $g$ and $\gm$ are not the same.
Suppose one seeks an inverse of the metric \ref{stpw} of the form
\be
\gm^{ij}=\al\dot{x}^i\dot{x}^j+\beta(\dot{x}^ix'^j+x'^i\dot{x}^j)+\ga x'^ix'^j+bg^{ij},
\label{guess}
\ee
then using \ref{stm} to form the identity,  dimension $d=\gm^{ij}\gm_{ij}$,
the $\beta$ and $\ga$ terms self-cancel and the $b$ term leads to a contradiction,
leaving the $\al$ term
\be
\gm^{ij}=-\fr{4\pi^2\al'^2d}{\ar^2}\dot{x}^i\dot{x}^j.
\label{show}
\ee
The inverse metric allows constuction of a Christoffell symbol which is a sum of
\ber
\st{1}{\Ga^i_{jk}}&=&\fr{\dot{x}^id}{\ar^2}
\left(\dot{x}\cdot x'x'_{(j,k)}+\dot{x}^lx'_jx'_{[l,k]}+\dot{x}^lx'_kx'_{[l,j]}\right)
,\lb{stch}\\
\st{2}{\Ga^i_{jk}}&=&\fr{\dot{x}^id}{2\ar^2}
\left(-\dot{x}^lx'^2\{_{jlk}\}-2\dot{x}_{(j}x'^2_{~,k)}+g_{jk}\dot{x}^lx'^2_{~,l}\right),
\nonumber
\ear
and from this a Riemann tensor can be constructed,  but its interpretation is obscure.
\section{The membrane.}\label{membrane}
The square root or non linear membrane action is
\be
S_D=k\int_M\rm{d}^{p+1}\xi\sq{-\ga},~~~
\ga_{ab}=g_{ij}\p_a x^i\p_b x^j,~~~
\sq{-\ga}=(-\det \ga_{ab})^\fr{1}{2},
\label{4.1}
\ee
so that
\be
\Lg=F(x,x^a)=k\sq{-\ga}.
\label{memF}
\ee
Choosing
\ber
p&=&1,~~~
k=-\fr{1}{2\pi\al'},~~~
a,b\dots=\ta,\si,~~~
\rm{d}^2\xi=\dt\ds,\\
\ga&=&\det(\ga_{ab})=-\ar^2,~~~
\ga_{ab}=\left(\begin{array}{cc}\dot{x}^2&\dot{x}\cdot x'\\
                                x'\cdot\dot{x}&x'^2\\
               \end{array}\right),~~~
\ga\ga^{ab}=\left(\begin{array}{cc}x'^2&-x'\cdot\dot{x}\\
                                   -\dot{x}\cdot x'&\dot{x}^2
               \end{array}\right),
\nonumber
\label{4.2}
\ear
the string lagrangian \ref{slag} is recovered.
The first fundamental form is defined by
\be
\ff^{ij}\equiv\ga^{ab}x^i_ax^j_b,~~~
\ff^{ik}\ff^j_{.k}=\ff^{ij},~~~
\ff^k_{.k}=\ga^c_{.c}=x^{k c}_{..}x_{k c}=p+1,
\lb{4.2b}
\ee
which allows the generalization of the projection tensors
\ref{ppproj} and \ref{3.1.16} to be expressed as
\be
h^{ij}=g^{ij}-\ff^{ij},~~~ h=d-1-p.
\lb{4.3}
\ee
One can form the Christoffel symbol for the metric $\ga$
\be
\{^a_{bc}\}=x^a\cdot x_{b,c},
\label{gach}
\ee
using the last equation of (\ref{4.2b})
partial derivatives can be replaced by covariant derivatives
\be
x_{;bc}=x_{,bc}-\{^e_{bc}\}x_e=x_{,bc}(1-x^ex_e)=-px_{,bc},
\label{gacd}
\ee
and \ref{gach} becomes
\be
\{^a_{bc}\}=-\fr{1}{p}x^a\cdot x_{;bc},
\label{gachcov}
\ee
which must be torsion free.
The Riemann tensor is defined by
\be
R^a_{~bcd}\equiv\{^a_{db}\}_{,c}-\{^a_{cb}\}_{,d}
+\{^a_{cf}\}\{^f_{db}\}-\{^a_{df}\}\{^f_{cb}\},
\label{defrie}
\ee
using \ref{gachcov} this is
\be
R^a_{~bcd}=-\fr{2}{p}\left(x^a\cdot x_{;[d|b|}\right)_{;c]}
+\fr{2}{p^2}x^e\cdot x_{;[c|b}x^a\cdot x_{;e|d]}
\label{riex}
\ee
contracting and using $0=(p+1)_{,e}=(x^ax_a)_{;e}=2x^ax_{;ae}$ gives
\be
R_{bd}=-\fr{1}{p}(x^a\cdot x_{;db})_{;a}+\fr{1}{p^2}x^e\cdot x_{;ab}x^a\cdot x_{;ed},~
R=-\fr{1}{p}(x^a\cdot x^d_{~d})_{;a}+\fr{1}{p^2}x^e\cdot x_{;ab}x^a\cdot x_{;e}^{~~b}.
\label{garic}
\ee
The Ricci scalar is a total derivative so that it has limited use as a classical lagrangian.
The lagrangian obeys a generalization of the homogeneity condition \ref{r1.8}
\be
F(x^i,k_Ax^{iA})=k_AF,
\label{memhom}
\ee
where $A$ is an unsummed internal index.
From \cite{mdrrm} the momentum and hessian are
\be
p^{i a}=\fr{\p\Lg}{\p x^{i a}}=+k\sq{-\ga}x^{i a},~~~
w^{ij ab}=\fr{\p^2\Lg}{\p x^{j b}\p x^{i a}}
             =+k\sq{-\ga}(g^{ij}\ga^{ab}+x^{i a}x^{j b}),
\label{2.3.2}
\ee
The generalization of the second order relationship (\ref{stf2}) is
\be
F^2_{x^{ia}}=2k\sqrt{-\ga}p^{ia},~~~
F^2_{x^{ia}x^{jb}}=-2k^2\ga(g^{ij}\ga^{ab}+2x^{bj}x^{ai})
\label{mtf2}
\ee
from which the generalization of the hat metric (\ref{stm}) is
\be
\gm^{ij}M^{ab}\equiv-\fr{1}{2}F^2_{x^{ia}x^{bj}}
=k^2\ga\left(g^{ij}\ga^{ab}+2x^{ia}x^{bj}\right),
\label{mtm}
\ee
where $M$ is a matrix to be determined.
There are three possible ways of removing the internal indices occuring in $M_{ab}$:
the {\it first} is to pick out a component
\be
\gm_{ij}M_{\tau\tau}=k^2\ga\left(g_{ij}\ga_{\tau\tau}+2\dot{x}_i\dot{x}_j\right)
\label{pos1}
\ee
proceeding as for (\ref{guess}) there are more possibilities,
but choosing that $\gm$ has only explicit $\dot{x}$ dependence and not $x'$ dependence
\be
\gm^{ij}=\fr{dM_{\tau\tau}\dot{x}^i\dot{x}^j}{k^2\ga\dot{x}^2(\ga_{\tau\tau}+2\dot{x}^2)},
\label{ipos1}
\ee
the {\it second} is to trace over the internal indices
\be
\gm_{ij}M^a_{~a}=k^2\ga\left((p+1)g_{ij}+2\ff_{ij}\right)
\label{pos2}
\ee
where the first fundamental form $\ff$ is given by (\ref{4.2b}),
the inverse metric is of the form
\be
\gm^{ij}=\alpha g^{ij}+\beta\ff^{ij},~~~
\label{pos2gen}
\ee
where $\alpha$ and $\beta$ are constrained by
\be
(d+2)\alpha+(p+3)\beta=\fr{dM^c_{~c}}{k^2\ga(p+1)}
\label{pos2constraint}
\ee
the {\it third} is to take determinants over the internal indices
\be
\gm_{ij}\det(M_{ab})=k^2(-\ga)\left(-g_{ij}-2\det_{ab}(x_{ia}x_{bj})\right)
\label{pos3}
\ee
where $\det_{ab}$ signifies that the determinant is taken over the internal indices $a,b$,
this choice does not seem to have an explicit inverse because of the determinant.
For (\ref{pos1}) and (\ref{pos2}) it is possible to construct Christoffel symbols
similar to (\ref{stch}).
\newpage
\section{Conclusion.}\label{conclusion}
The string and membrane generalization of the finslerian point particle approach to riemannian
geometry was presented.   The relationship between the hat metric $\gm$ and the indicial
metric $g$ given by (\ref{stm}) and (\ref{mtm}) is no longer an equality as terms dependent
on the internal properties of the string or membrane appear.
This macroscopic dependence on microscopic internal properties might be small enough
to produce realistic models but large enough to lead to new predictions.
There is the possibility that this geometric picture might have a thermodynamic analogy,
compare \cite{jacobson},  in which entropy could be assigned to the
string's area $\ar$ and related to macroscopic properties.   The geometry used does not involve
$\hbar$ so that the relationship between microsopic and macroscopic is classical:  a more
usual picture would be to take it that quantum and many body properties of strings are
necessary to build medium size systems,  in other words that physicaly intermediate lenght
scale properties are necessary.   Alternatively the geometry presented here could be interpreted
as that the geometry of microscopic spacetime.
\section{Acknowledgements}\label{acknowledgements}
I would like to thank David Bao,  Howard E. Brandt and Lajos Tamassy for correspondence in
september 2000,  and  J\"urgen Jost for conversations in 2007 on geometry based on area.

\end{document}